\documentclass[10pt,conference,letterpaper]{IEEEtran}

\usepackage{amsmath,amssymb,array,color,colortbl,graphicx,multirow}
\usepackage{microtype}
\usepackage[pdfpagelabels=false]{hyperref}
\usepackage{comment}

\usepackage{algorithm2e}
\usepackage{comment}

\usepackage{framed,color}

\definecolor{mygreen}{rgb}{0,0.5,0}

\newcommand{\stefan}[1]{\textit{\textcolor{red}{[stefan]: #1}}} 
\newcommand{\system}{RVaaS}

\begin{document}

\title{Discreet Client Isolation Verification for carrier Networks}

\title{Relaxing the ISP Trust Assumptions:\\
Confidential Verification of Network Routes}

\title{Verifiable Routing Despite Untrusted ISPs}

\title{Verifiable Routing in the Presence of Untrusted ISPs}

\title{Verifiable Routing With Untrusted Network Operators}

\title{The Untrusted Network Operator Problem}

\title{Integrity Checks for Insecure Network as a Service}

\title{{\Large Routing-Verification-as-a-Service~(RVaaS):}\\Trustworthy Routing Despite
Untrusted Operators}

\title{{\Large Routing-Verification-as-a-Service~(RVaaS):}\\Trustworthy Routing in the Era of\\ Compromised Network Management Systems}

\title{{\Large Routing-Verification-as-a-Service~(RVaaS):}\\Trustworthy Routing Despite Cyber Attacks}

\title{{\Large Routing-Verification-as-a-Service~(RVaaS):}\\Trustworthy Routing 
Despite Insecure Providers}

\author{Anonymous}

\author{}

\author{
Liron Schiff$^1$  \quad Stefan Schmid$^{2}$\\
{\small~$^1$ Tel Aviv University, Israel \quad~$^2$ Aalborg University, Denmark \& TU Berlin, Germany}
}

\author{
Liron Schiff$^1$ \quad Kashyap Thimmaraju$^{2}$ \quad Stefan Schmid$^{3}$\\
{\small~$^1$ Tel Aviv University, Israel \quad~$^2$ TU Berlin \& T-Labs, Berlin, Germany \quad~$^3$ Aalborg University, Denmark \& TU Berlin, Germany}
}

\date{}

\maketitle \thispagestyle{empty}

\sloppy

\begin{abstract}
Computer networks today typically
do not provide
any mechanisms to the users to learn,
in a reliable manner, 
which paths have~(and have not!) been taken
by their packets.
Rather, it seems inevitable that as soon as 
a packet leaves
the network card, the user is forced to trust
the network provider 
to forward the packets as expected or agreed upon.
This can be undesirable, especially in the light of
today's trend toward more programmable
networks: after a successful cyber attack on the network
management system or Software-Defined Network~(SDN) 
control plane, an adversary in principle has complete control over the network. 

This paper presents a low-cost and efficient solution to 
detect misbehaviors and ensure 
trustworthy routing over untrusted or insecure providers, in particular
providers  
whose management system
or control plane has been compromised (e.g., using a cyber attack).
We propose Routing-Verification-as-a-Service~(\system): 
\system~offers clients a flexible interface to 
query information relevant to their traffic, while respecting 
the autonomy of the network provider.
\system~leverages key features of OpenFlow-based SDNs
to combine (passive and active) configuration monitoring,
logical data plane verification 
and actual in-band tests,
in a novel manner.
\end{abstract}

\section{Introduction}\label{sec:intro}

While improving the security of the Internet routing system has
been a prime concern for many years already,
 the interface between Internet users
(including companies)
and the network provider~(e.g., the carrier or datacenter operator)
has received little attention.
Today, the user typically does
not even have any means to \emph{specify} desired 
and undesired routing paths~(e.g., using
white or black lists), and even less is supported
in terms of \emph{verification}.
Rather, it is often implicitly assumed that 
the user needs to trust its network provider, including its
network management system software, unconditionally.

While 
traceroute and trajectory sampling tools may be sufficient 
to verify routes in 
regular networks~\cite{bck-ts,ts}, and
may still perform well in the context of faulty 
and heterogeneous networks~\cite{unrel-ts,csamp},
they are insufficient in non-cooperative and adversarial
environments: an unreliable network operator may simply
not reply with the correct information, also breaking any scheme based on packet
labeling or tagging~\cite{pathquery,pathtracer}.
Even more challenging than verifying 
the used paths, is to \emph{test avoidance}, i.e.,
verifying that certain paths have not been taken
and certain destinations have not been reached~\cite{alibi-routing}.

The threats introduced by untrusted providers are manifold. 
In particular, routing can be compromised even in scenarios where the 
provider itself is in principle benign. 
For example, over the last years, numerous flaws have been found
in network management systems~\cite{flaw}.
The problem is exarcerbated in Software-Defined Networks~(SDNs)~\cite{redhat}:
A new SDN control plane may be vulnerable to cyber attacks,
which, given the important role the SDN controller plays compared to more distributed
legacy network protocols, is particularly worrying: an adversary with access to the
control plane can in principle arbitrarily change the network forwarding behavior,
and violate security policies~(e.g., breaking logical isolation domains between
health care providers~\cite{header-space}) or exfiltrate
confidential traffic. Today,
clients do not have a means to reliably verify the
data plane configuration. 
 
At first sight, the problem seems to be an inherent one: 
as soon as the packet enters the provider network,
its fate is inevitably decided by the provider and its
network management system and software control plane. 
While 
a~(possibly signed) acknowledgment from the receiver may eventually confirm
to the sender
that the packet has successfully arrived, this is insufficient as it does not
provide any information about which paths have been
taken and which~(possibly \emph{additional}) destinations have been
reached. 
The problem is particularly cumbersome in the context of 
high-performance networks where cryptographic per-packet operations 
(like encryption, signatures, etc.) are out of question.

This paper is motivated by the question whether
it is possible to reduce the seemingly inevitable
trust assumptions in the network provider, and to empower the user
to verify the routes taken and destinations reached by its packets.
Ideally, the resulting solution should also 
not introduce significant computational overheads,
and also respect the autonomy of the network operator:
security and business critical details of the underlying topology should not be revealed.

\subsection{Our Contributions}

This paper presents 
\emph{Routing-Verification-as-a-Service~(\system)}, 
a novel network service which allows users~(or more generally:
\emph{clients}) to query and verify
relevant properties of the network routes installed on their behalf.
\system~removes
the need for users to unconditionally trust 
the network providers to forward their packets
according to the agreed upon routing
policies, and also accounts for the possibility
that operators or control software is compromised,
e.g., due to a cyber attack. 

\system~is based on passively and actively monitoring network configurations,
and on the in-band interception of user request messages
(e.g., using OpenFlow \emph{Packet-ins}). Upon a query request,
\system~performes a static packet trajectory analysis
(identifying relevant endpoints),
and actively issues verfication packets and client authentication
tests (e.g., the verify that endpoints are legitimate).

\system~features the following
properties:

\begin{enumerate}

\item \textbf{Verifiable routing properties:} Users can 
learn about and verify, through a flexible interface,
relevant information related to the routes taken by their packets,
such as \emph{the set of destinations}, or whether \emph{fairness conditions}
are fulfilled~(e.g., regarding bandwidth allocations). 
For example, users can verify that their traffic is not routed
in a way which violates privacy, e.g., is not exfiltrated
or routed through certain geographic regions.

\item \textbf{Confidentiality:} The autonomy of the provider
is preserved, and 
security or business critical topological details can be kept confidential.

\end{enumerate}

One attractive feature of our approach is that it allows
users  to issue
very general queries, which are not limited to connectivity
alone, but may also include geographic, performance and fairness aspects. 
Another feature is the provided \emph{modularity}:
queries may not be limited to a single provider but may
recursively span consecutive networks along a route.

To provide the \system~service, it is sufficient
to deploy a single secure server,
somewhere in the network; additional~(independent)
servers can increase the security further.
These servers do not have to inspect live traffic, and have low resource requirements;
they also does not come with strict latency requirements. 

\subsection{Paper Scope and Novelty}

We emphasize that the goal of our approach is
to empower the users to \emph{detect} misbehavior, 
as opposed
to \emph{prevent} misbehavior. In other words, alone, our approach
is unable
to ensure a user's packets will not traverse certain
network regions or reach certain destinations.  However, 
we believe that the possibility
to detect misbehavior can often be a strong disincentive 
to deviate from the correct behavior.
Moreover, we in this paper do not consider the orthogonal
question
of how a user should specify its desired and undesired
routes to the network provider. 

Generally, we believe that our work assumes an interesting 
position in the secure
routing space. While there has
been much interest in 
securing the inter-domain routing protocol
or in dealing with unreliable
data plane components, 
we study how to reduce trust
assumptions in the entity installing
the rules on the routers in a single administrative domain.
Moreover, we make the case for marrying
verification mechanisms in the ``logical space''~(e.g., \emph{which routes
exist?}), with physical verification 
mechanisms in the data plane~(e.g., \emph{which
host destinations are actually reached?}). 

We also note that while for $\system$, any
secure server is in principle sufficient, 
our architecture can also benefit from the advent of 
novel hardware developed in the context of
\emph{Intel SGX}~\cite{sgxr3,sgxr1,sgxr2,white}. In this respect,
we see our work also as an interesting case study demonstrating
a new application of this technology in the context of
secure routing.

\subsection{Organzation}

The remainder of this paper is organized as follows.
Section~\ref{sec:background} provides
necessary background on SDN/OpenFlow. 
Section~\ref{sec:model} introduces our model together
with some terminology. 
RVaaS is described in detail in 
Section~\ref{sec:solution}. 
After reviewing related work in Section~\ref{sec:relwork},
we conclude our contribution in 
Section~\ref{sec:conclusion}.

\section{Background}\label{sec:background}

The solution proposed in this paper is tailored
for 
Software-Defined Networks~(SDNs), and
we will provide the necessary background accordingly
in this section. 

In a nutshell, Software-Defined Networks~(SDNs)
outsource and consolidate the control over
the data plane devices~(the switches or routers)
to a logically centralized software controller.
This decoupling introduces flexibilities and
innovation opportunities, as the control plane
can now evolve independently from the constraints
of the data plane~\cite{road}. 
OpenFlow is the de~facto SDN protocol standard today.
OpenFlow 
is based on a match-action concept: OpenFlow switches store
rules a.k.a.~flow entries~(installed by the controller) consisting of a match and an
action part. A packet matched by a certain rule will be subject
to the associated action.
For example, an action can define a port to which the
matched packet should be forwarded, or add or change a tag
(a certain part in the packet header).
In OpenFlow networks, the distinction between switches and routers 
disappears: an OpenFlow switch can match~(and apply actions to) 
not only layer-2 but also layer-3 
and layer-4 header fields.

An OpenFlow switch can~(and should) be connected to one or
multiple controllers
via an authenticated and secure communication channel 
(e.g., SSL/TLS).\footnote{We note however that according to a 2013 study, only 
2 out of
8 OpenFlow switches and 1 out of 8~(popular) OpenFlow controllers fully support.~\cite{kevin}} 
Thus, only legitimate controllers can send rule updates to the switch.

In order for the controller to learn about newly arriving flows, 
OpenFlow switches can forward packets to the controller by 
sending them within so-called \emph{Packet-In} messages. 
Similarly other events~(link failures, switch errors, etc.) are reported 
as well to the controller using dedicated OpenFlow messages. 

As a reaction to such events, a controller may want to
change the installed flow on the switch~(using
\emph{Flow-Mod} commands) or explicitly send packets out from the switch~(using \emph{Packet-Out}).
Moreover, 
to stay informed about the current configuration of a switch~(the existing
flow entries), the controller should use the 
OpenFlow \emph{add flow monitor command}.

\section{Model and Threat}\label{sec:model}

We consider a software-defined network servicing multiple 
clients which are geo-spatially distributed. The client and provider roles are defined as follows:
\begin{enumerate}
\item \textbf{The Clients~(or Users):}
We will refer to the users or communication endpoints of the network
as the \emph{clients}.
Each client may be connected to the network infrastructure 
at multiple access points~(switch ports), and request connectivity and
routing services~(regarding his access points) from the provider.

\item \textbf{The Provider:} The provider running the software-defined network 
consists of two parts:
\begin{enumerate}
\item \emph{Network management system and control plane:}
A software in charge of defining and installing the device configuration
(e.g., routing policies), within the constraints defined by the
clients. 

 \item \emph{Infrastructure:} Routers (resp.~OpenFlow switches) and links.
\end{enumerate}
 \end{enumerate} 

We consider a threat model motivated by \emph{cyber attacks}: 
an external attacker which compromised the network management or control plane 
(e.g., using a Trojan or a remote cyber attack) 
aims to change the data plane configuration, e.g.,
to divert client 
traffic to unsupervised access points or through undesired jurisdiction, 
thereby putting the security of the network and the traffic privacy at risk.
However, while the network management system and control plane may be hacked,
we assume the infrastructure to be secure:
The question of what security properties can be guaranteed
in scenarios where control planes can be compromised and malicious 
while the data plane is correct, is scientifically interesting on its own right.
However, 
we argue that the question is also a practically relevant one, in three respects:
\begin{itemize}
\item
While a cyber attacker~(not an insider!) may be able to hack
the management and control plane, it is impossible to change
physical configurations from remote locations.

\item In the context of network virtualization, the physical
infrastructure provider and the virtual network operator
are often considered two different roles. In this respect,
our model can be understood as a case study of 
how to deal with a malicious virtual network operator.

\item Our model can also be motivated by the current trend
toward more trusted hardware (see, e.g., Intel SGX).
\end{itemize}

The clients can also be untrusted in our model, and may
for example not inform the sender about having
received packets, or may try to infer confidential
details about the network topology.

Our objective is  to
enable a trustworthy routing,
by empowering a client to find out
about and verify relevant properties of the routing
applied to its packets~(e.g., the set of
reached destinations). 
Moreover, the autonomy of the provider should be preserved.
In particular, clients should not be able to infer the topology or
critical features~(like bottlenecks) of the network
itself.
Finally, details of the client should not leak: 
the provider should not learn about their queries
(whose content is somewhat confidential). 

In general, we assume 
a high-speed network~(e.g., Internet backbone
or datacenter), where 
per-packet encryptions or
public key operations are hardly used due to the high costs of deploying 
and maintaining them. Concretely, we rule out
signed logs in every packet, per-flow state in
forwarders~(which stymies fail-over), and ideally not even
per-flow public key operations.

In summary:
\begin{itemize}
 \item Switches are trusted~(e.g., bought from a trusted vendor),
 and are initially configured correctly. 
 \item Internal network ports are known, and follow a well-defined wiring plan.
 \item Links are trusted: no physical taps are installed.
 \item Switch to \system~controller sessions are secured, 
 using encrypted  OpenFlow sessions and apriori configured switch certificates for authentication.
 \end{itemize}

\section{Trustworthy Routing}\label{sec:solution}

We will first discuss the main ideas and concepts
behind Routing-Verification-as-a-Service~(\system).
Subsequently, we give an example and 
discuss extensions
and limitations.

\subsection{Main Concepts}

At the heart of \system~lies 
a flexible interface which allows the 
clients to query relevant information 
related to how their packets are being forwarded in the network.
The interface 
allows clients to ask questions such as:
\begin{itemize}
\item Which destinations~(resp.~other clients and hosts) can be reached
by the traffic
leaving my network card? This question may also be made more specific,  
e.g.,
constrained to traffic within a certain header space.

 \item For which sources~(e.g., other clients, hosts) currently 
 exist routing paths which can reach my network card? Again,
 the question can be made more specific.
 
 \item Is my traffic forwarded fairly, e.g., according to network neutrality principles?
\end{itemize}

Generally, queries related to connectivity, path lengths optimality, 
traversed geographic regions, traffic shaping, 
quality-of-service~(e.g., dedicated bandwith), 
etc.~are 
supported.
A client may also request a compact representation of the transfer 
function of its offered routing service. 

Through attestation, the client can verify that $\system$ 
is the one that securely responds to its queries. Moreover, 
the provider makes sure that the correct $\system$ application
is operating on the server, and not a fake one that may leak sensitive information regarding the infrastructure or clients.

\system~is based on a passive-active approach: 
events in the data plane are monitored, and analyzed in the control plane; 
upon a client request, endpoints are actively tested and authenticated.
\system~can be realized using a stand-alone OpenFlow controller,
henceforth called \emph{\system~controller},
which monitors the configurations of all the switches, and 
which may send and receive~(resp.~intercept and inject) specific messages 
in-band, in order to communicate with the clients.
This controller is different from possibly additional controllers used
by the network provider to manage the network,
and should be trusted. Also, while a single one is in principle enough, 
different entities (e.g., a certification authority) 
may provide different independent controllers,
reducing the attack surface further.

In order to provide its service, the \system~controller performs
three different functions: passive or active configuration monitoring,
logical data plane verification, and actual in-band testing using client interaction.

\subsubsection{Configuration Monitoring}

Our stand-alone \system~controller
is secured (i.e., installed correctly, and cannot be influenced by the network provider)
and connected to all switches,
via authenticated and encrypted OpenFlow sessions.
Through these sessions, the controller maintains an up-to-date
snapshot of the network configuration, either passively
(monitoring events) or actively (query the switch state 
or issue~and later intercept LLDP like packets through all internal ports). 
This information is acquired in a manner
that cannot be exploited by an untrusted operator.

\begin{figure}[t]
    \begin{center}
        \includegraphics[width=0.99\columnwidth]{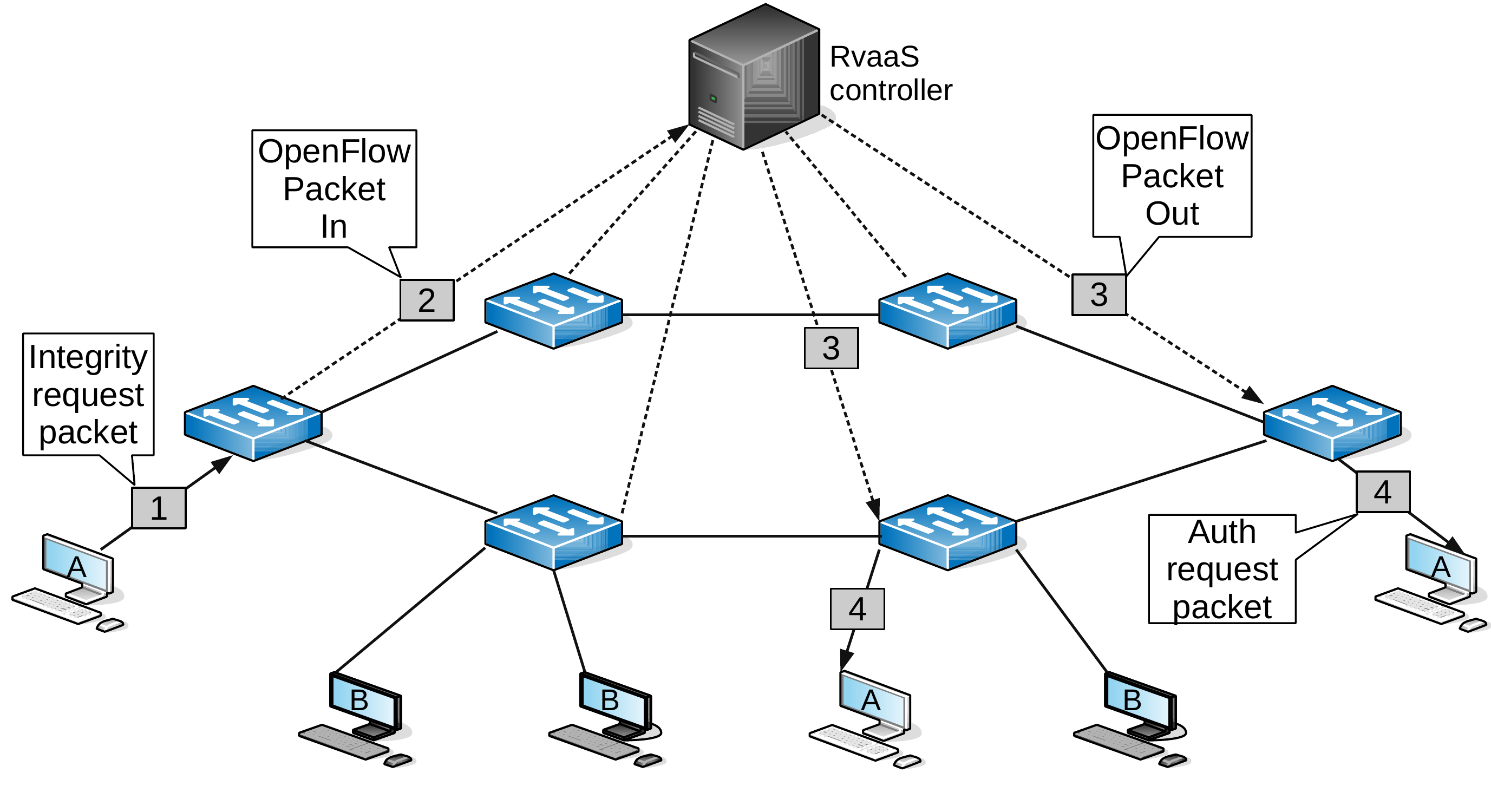}\\
    \end{center}
    \caption{A client makes an integrity request to the \system~controller. The \system~controller analyzes the request and then dispatches Auth(entication) request packets to relevant
clients.}
    \label{fig:request}
\end{figure}

\subsubsection{Logical Verification}

In order to answer client queries,
relevant routes are computed in the logical space, given the current
network snapshot collected by the \system~controller.
For example, the \system~controller may
perform Header Space Analysis~\cite{header-space},
or simply emulate the network based on the current configuration.

\subsubsection{In-band Test \& Client Interaction}
The \system~controller
has active access to the network traffic in order 
to collect user requests/queries. Such client messages have distinct properties 
(e.g., destination address, VLAN tag, etc.) that allow them 
to be matched at the~(ingress) switches and reported to the controller. 
In addition each client knows the public key
of the \system~controller, allowing it to encrypt messages and 
verify authenticity of the results.

In some cases, answering user requests, 
involves sending  further requests
(e.g., using OpenFlow packet-out commands). For example, these packets trigger
destination clients to respond to the querying clients, in an authenticated manner
(\emph{authentication requests}).
Towards this end, clients run a software which responds 
to our authentication requests, in user space, publishing
themselves by sending a UDP packet (with a specific magic header field value
which can be intercepted and traced back to the origin, due to the logically
centralized view). 
Concretely, in order to not expose $\system$ to cyber attacks
itself, our solution is based on in-band interception (namely OpenFlow packet-ins) 
of user request  messages (e.g., using a magic header value); responses
are sent via packet-outs. 
That is, $\system$ is only reachable via a very simple OpenFlow interface and indirectly;
no special protocols and servers are needed. 

In general, the security of our architecture relies on the integrity
of the secure hardware, as well as on the frequency of
the network snapshot it takes.
In particular, $\system$ needs to
ensure that it receives all the relevant updates from the switches.
This is guaranteed in our setting where OpenFlow switches
are reliable. However, additionally, it is also possible for
$\system$ to proactively query the switches for their current
configuration. The latter however needs to happen at random
times, which are hard to guess for the adversary. This is important
as otherwise, the adversary may simply set the correct rules
for the short time periods in which the box checks the configuration.
Short term reconfiguration attacks can also be prevented
by maintaining some history. 
Regarding the confidentiality of the network topology
and respecting the provider's autonomy, our
architecture offers many flexibilities. In particular,
queries can be limited to learn only about 
endpoints, but nothing about the actual
routing paths inside the network.

\subsection{Case Studies}

\begin{figure}[t]
    \begin{center}
        \includegraphics[width=0.99\columnwidth]{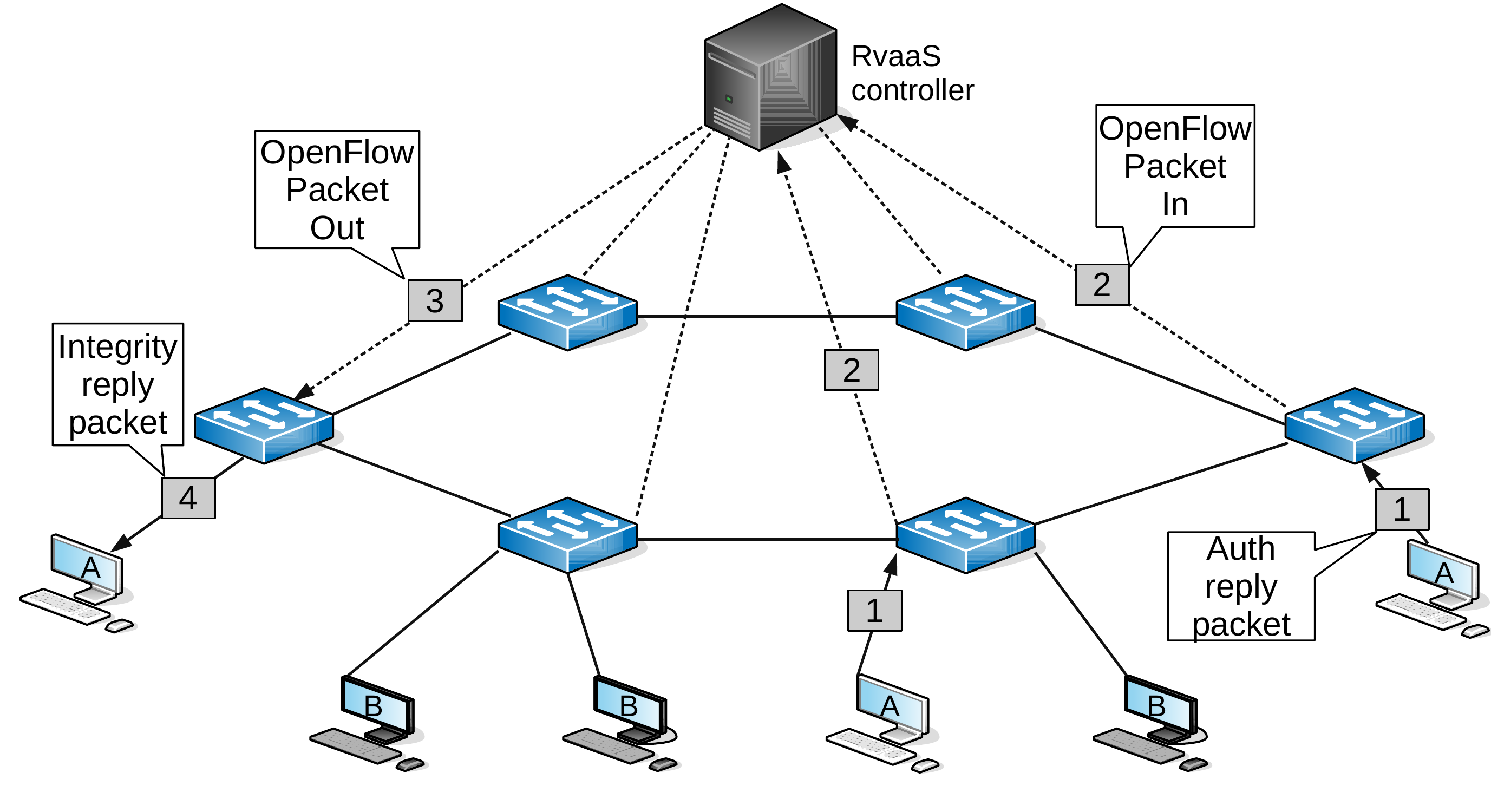}\\
    \end{center}
    \caption{Relevant clients send Auth(entication) reply packets back to the \system~controller which collects the replies and sends them back to the requesting client.}
    \label{fig:response}
\end{figure}

Let us consider some relevant case studies.
We refer the reader to Figures~\ref{fig:request}
and ~\ref{fig:response} for
some illustrations.
 
\subsubsection{Isolation Checks}

One of the most fundamental security queries
supported by \system~regards whether the sub-networks 
where different clients are located, are isolated from each other:
no client can gain access to another client's network except 
through some access points used by the client. Failing to 
guarantee such a requirement makes the client vulnerable to 
join attacks in which an attacker first 
manipulates the network operation, and secretly adds 
access points which can then be used 
to access and/or damage client assets 
(such as private data  or hardware) managed by the network.

Our system can detect violations of isolation as follows: 
a client request, sent through an access point~(the request point), 
is intercepted at an ingress switch and reported through 
a \emph{Packet-In} to the system server. Based on its current network view,
the server computes all the possible access points that can communicate 
with the request point, e.g.,
using reachability tests based on Header Space Analysis. 
Given these access points, the server issues a \emph{Packet-Out} 
message at the corresponding outgress
ports of the network. The hosts behind these ports respond, with  
authenticated messages, which are intercepted and reported to the server. 
The collection of these authenticated
messages are then forwarded to the querying client, which
can verify the correctness and authenticity of these destinations. 

Note that the server also forwards to the client 
the total number of authentication requests that were made,
such that it can detect cases where some access points did not 
respond. 

\subsubsection{Geo-Location Checks}

As a second case study, we present a query made by a 
client to discover the locations where its traffic passes through. 
This is relevant, e.g., in scenarios where 
different jurisdictions exercise different privacy policies regarding user data.
First we require that the locations of all the switches 
(and preferably also the links) are known to the \system~controller. 
These locations can be revealed/estimated in each of the following ways:
(1) either disclosed by the infrastructure provider; (2) 
 collected from the clients themselves in a crowd-sourcing manner: 
clients can e.g., report their geographical locations which allows $\system$
to guess the location of nearby switches; 
 (3) or passively inferred from clients traffic, e.g., using geo-IP mappings, domain name records information, time zone estimations, etc.
 
Given a client geo-location request, the \system~controller 
uses header space analysis 
to find out all the intermediate and end point switch~(and link) 
of any possible route for the client. 
Then using  the locations of the network provider
components, the set of locations exposed to the client traffic is 
computed and sent to the client.

\subsection{Extensions and Discussion}\label{sec:discussion}

This section provides a discussion of our approach,
and identifies limitations and extensions.

\paragraph{Multi-Provider Settings}
While we have described our architecture for a single-provider
setting, in principle, our approach can also be used
across multiple providers. In this case, queries
need to be propagated between the $\system$ servers
of the respective providers. Clearly,
the trust assumptions then need to be extended
accordingly, to those servers as well.

\paragraph{Supported Queries}
In principle, a wide range of queries can be supported
within our framework, beyond simply identifying 
reachable clients. For example, given the up-to-date
network view, performance
and fairness related queries by clients may
be answered. Moreover, $\system$ could be used
to check
whether allocated routes and meter tables
meet network neutrality requirements.
Moreover, a slightly more complex service may also
maintain some history of the recent past, 
allowing $\system$ for example to traceback
the ingress port of an attack.

\section{Related Work}\label{sec:relwork}

Our paper assumes an interesting
new position in the secure routing space.
Arguably the most intensively-studied problem in the 
secure routing literature
regards how to ensure 
authenticity and 
correctness of topology propagation and route computation
across multiple untrusted and insecure domains,
e.g., by extending~\cite{sbgp,psbgp}
or redesigning~\cite{OPT,SICON}
the Border
Gateway Protocol~(BGP).
Moreover, the problem of how to design
secure routing protocols which
allow to deal with untrusted and insecure
switches and routers
currently 
experiences a rennaissance~\cite{untrusted-router-model,icing,perlman}.
In contrast, we in this paper
investigate mechanisms which empower \emph{the user}
to deal with untrusted or insecure operators, subject
to a cyber attack (from an external adversary without
physical access).
Our problem is also different from the recently
introduced and interesting malicious administrator
problem~\cite{matsumoto2014fleet,swste16}:
in that problem, it is assumed that a network is redundantly managed by 
multiple administrators or controllers, out of which only a minority
can be malicious. This allows for simple (yet crypto intensive)
secure solutions based on threshold objects and majority decisions.
In the context of operator networks, such a redundancy is not available,
and to the best of our knowledge, the threats introduced
by a malicious network operator have not been studied before.

We are not the first
to identify new security-related opportunities and challenges introduced
by the software-defined networking paradigm~\cite{redhat,Kreutz,sdn-sec-dc},
 also 
regarding traffic
monitoring~\cite{netsight,FleXam,Yu:2014:DCT:2620728.2620739}.
While the static logic of $\system$ can be implemented using 
Header Space Analysis (HSA)~\cite{header-space},
over the last years many alternative tools have emerged~\cite{veriflow,anteater,flowvisor}.
These tools in turn rely on early works on reachability 
by Xie et al.~\cite{xie},
and are not limited to switches and routers but 
can also be employed, e.g., in the context of firewalls~\cite{firmato,bgp-stat,fang,FIREMAN}.
Our work is orthogonal in the sense that \system~can benefit from such systems
in order to implement its query interface, while performing
the required authentication requests in the data plane.
That is, \system~in some sense combines 
data plane~\cite{chimera,gigascope,frenetic,monocle}
and
control plane~\cite{decor,pier,netquery} query systems,
and issues a minimal amount of requests in the data plane
(e.g., to collect information about attached clients).
 In this sense, our work is also orthogonal
 to recent literature
aiming to improve the latency at which network monitoring information 
can be retrieved~\cite{planck}.

\section{Discussion and Conclusion}\label{sec:conclusion}

This paper initiated the study of 
trustworthy routing architectures in the
context of hacked and untrusted network management
systems and control planes, as well as malicious
virtual network operators.
We have identified different requirements
and different roles in such a setting,
and provided a first solution which,
based on a secure but simple hardware, 
allows to decouple the roles and
empower clients to verify routing properties
while preserving the autonomy of the
operator, e.g., by respecting the confidentiality
of topological details.

While the underlying concepts may be more general, $\system$ 
is particularly well-suited for SDNs based on OpenFlow:
OpenFlow's match-action interface
provides an ideal technological basis for our approach.
In particular, OpenFlow enables a simplified 
monitoring of different equipment, the 
interception and injection of packets in order to communicate with the clients, 
without affecting existing services, and the 
centralized view and analysis of the collected configuration.

Clearly, at least initially, $\system$ targets power users, but
in the longer run, may also be incorporated into security/privacy products 
directly, and made available to end users. 
Our work also raises the question why an operator
would be willing to install $\system$. Besides the possibility
to consolidate logical network view and physical configuration
(e.g., in scenarios where the operators does not necessarily 
trust the SDN controller software to be perfectly correct),
we also see economic incentives: a telco hosting one or multiple
(independent) $\system$ servers may appear to be more trustworthy
to their customers, which can constitute a business advantage.
For instance, customers relying on security-critical networks, such as governmental networks,
are likely to prefer certified telcos, which offer an independent means
of verification.  

In general, we believe that our work assumes an interesting
new perspective on the classic topic of secure routing,
in several respects. For example, we believe that our distinction of network operator 
from physical infrastructure provider is an interesting and timely one, beyond the
considered cyber attack threat model: 
in the context of network virtualization and with the ongoing infrastructure
liberalization trend, network operators are more and more seen as
a business role which may be independent from the infrastructure owner.
Moreover, while our assumption of trusted infrastructure is a strong one, 
we believe that it constitutes more than an academic exercise: given today's
trend toward trusted hardware, our work is timely and provides an interesting
new look on this trend from a networking perspective. 

We understand our work as a first step.
In particular, while we show the potential for a more
trusted routing in less trusted environments, 
much more research is required to understand
the minimal assumptions required to implement such an
architecture, as well as to understand the fundamental
tradeoffs in terms of security and performance. 
It is also clear that there are inherent
limitations to such a solution. 
For example, it seems impossible to
deal with untrusted network operators who also
have physical access to the network, at least
in the classical, non-quantum physics world.

\section*{Acknowledgments}
Research supported by the German Federal Office for Information Security (BSI).
In particular, the authors would like to thank Jens Sieberg.

{\footnotesize
\def\noopsort#1{} \def\No{\kern-.25em\lower.2ex\hbox{\char'27}}
  \def\no#1{\relax} \def\http#1{{\\{\small\tt
  http://www-litp.ibp.fr:80/{$\sim$}#1}}}\def\noopsort#1{}
  \def\No{\kern-.25em\lower.2ex\hbox{\char'27}} \def\no#1{\relax}
  \def\http#1{{\\{\small\tt http://www-litp.ibp.fr:80/{$\sim$}#1}}}

}

\end{document}